# A Framework for Validating IS Research Based on a Pluralist Account of Truth and Correctness

**John Mingers (Corresponding author)**
Kent Business School, University of Kent, Canterbury CT7 2PE, UK
j.mingers@kent.ac.uk
01227 824008

**Craig Standing**
Dept. of Business and Law, Edith Cowan University, Perth, Australia
c.standing@ecu.edu.au

**Abstract**

Research in information systems includes a wide range of approaches which make a contribution in terms of knowledge, understanding, or practical developments. The measure of any research is, ultimately, its validity – are its finding true, or its recommendations correct? However, empirical studies show that discussion of validity in research is often weak. In this paper we examine the nature of truth and correctness in order to construct a validation framework that can encompass all the varied forms of research. Within philosophy, there has been much debate about truth – is it correspondence, coherence, consensual or pragmatic – and in fact current views revolve around the idea of a pluralist view of truth – it is one and many. Related to truth is the concept of correctness, and in particular the necessity of both internal correctness and external correctness. The framework we develop based on these concepts of truth and correctness has been applied to a range of research forms including positivist, mathematical, interpretive, design science, critical and action oriented. The benefits are: i) that a greater and more explicit focus on validity criteria will produce better research; ii) having a single framework can unite what at times seem conflicting approaches to research; iii) having criteria made explicit should encourage debate and further development.

**Key Words**: action research, correctness, critical research, design science, interpretive research, positivist research, simulation, truth, validation

# A Framework for Validating IS Research Based on a Pluralist Account of Truth and Correctness

## INTRODUCTION

Information systems is a wide ranging discipline involving varied forms of research with different purposes. There is research aimed at producing knowledge, from a variety of perspectives – positivist (Dubé et al. 2003; Straub et al. 2004), interpretive (Klein et al. 1999; Walsham 2006a), critical (Klein et al. 2004; Mingers 2004) and more; research that aims at producing software or IT/IS artifacts – design science (Hevner et al. 2004); and research that hopes to bring about improvements to organizational problems – action research (Chiasson et al. 2009). These heterogeneous forms of research are carried out in many different ways; based on different and sometimes conflicting assumptions; and often use fundamental concepts such as "information", "theory", "causality" or "knowledge" incompatibly.

We are not against the idea of pluralism in IS research at all (Mingers 2001a; Mingers 2001c) but we do agree with Lee (Lee 1991; Lee et al. 2009; Lee et al. 2014; Lee et al. 2015a) that there needs to be some degree of coherence or rigor underlying these multifarious approaches in order to justify and validate the results that end up being published in our journals and used as a basis for affecting peoples' lives.

In this paper we will investigate one, crucial, element of research – that of truth or correctness which in many ways underlies all the others. Scholarly research in any field aims to produce knowledge, not least within IS where we also have the specific domain of knowledge management which appears to have knowledge as its subject matter (Mingers 2008). This immediately begs the question of what exactly is knowledge, and how does it differ from mere belief or opinion? Traditionally, within philosophy, knowledge is said to be "justified true belief" (Gettier 1963; Pritchard 2006), that is, it is a kind of belief or opinion but one for which we have evidence or warrant, and, essentially, is actually true whether or not we can in fact determine its truth1. This leads to the further question, what exactly is truth for unless we know what truth is, we cannot understand what knowledge is2

That truth is indeed a goal of IS research has been expressed, for example, by Straub et al (2004) "The purpose of validation is to give researchers, their peers, and society as a whole a high degree of confidence that positivist methods being selected are useful in the quest for scientific truth" (p. 383). However, in most papers, including that one, the actual nature of truth and how it might be discovered is little discussed.

There is a traditional view within philosophy - the correspondence theory of truth (Lynch 2001) – perhaps best expressed by Aristotle: " To say that that which is, is, and that that which is not, is not, is true". To some extent this is a truism, but to clearly articulate a theory of truth we need to specify its elements: what is it that can have this truth property (the truth bearer); what is it that could make the truth bearer true (the truth maker); and what is the nature of the correspondence relation?

There have, however, been many criticisms of the correspondence view of truth, particularly in terms of its realist view of the external world, and this led to a number of alternatives. For example, coherence theory which evaluates a belief in terms of it coherence or consistency with other well-attested beliefs (Walker 1989); pragmatism which focusses on long-term success in practice (James 1976; Peirce 1878) or consensus theory (Habermas 1978) which sees truth as that which a relevant community of enquirers agrees about. A more radical approach, known as deflationism (Quine 1992; Strawson 1950), suggests that actually truth has no substantive nature to be explained, and that it is really just a linguistic pseudo-problem.

---

[1] There are very different views of knowledge, for example Foucault (1980) who sees knowledge as ultimately constituted through power, and postmodernism that perhaps denies the possibility of knowledge at all but these will not be pursued in this paper.
[2] We recognize the inevitable circularity here – knowledge requires truth, but truth requires knowledge.

In the face of the seeming stand-off between these competing positions, a new approach has been developing that aims to retain the idea that truth is a substantive concept, and some form of realism about the relation to the external world, whilst accepting the criticisms of standard correspondence theory. This approach involves a pluralist view of truth – truth is one and truth is many – there are generic characteristics of truth but these may be realized differently in different domains (Lynch 1998; Pedersen et al. 2013c). In the physical domain one might hold a correspondence view while in the mathematical domain one might have a coherence view.

Whilst truth may be a defining characteristic of knowledge, as we saw above not all IS research aims purely at knowledge - design science aims to produce effective software or artifacts, and action research aims to solve problems in organizations. In these domains it may not be appropriate to talk about truth but rather the related term *correctness* (Engel 2013; Thomson 2008). It seems more appropriate to say a computer system works" correctly" rather than "truly". In many areas they seem equivalent – if a belief is true then it will also be correct, while a belief that is incorrect would thereby be false. But correctness is a wider term than truth in that it applies to things other than beliefs or propositions, for example actions or procedures.

The purpose of this paper, therefore, is to develop a general conceptualization of truth and correctness that can be applied across all areas of research in IS. Essentially, this will specify criteria for evaluating the rigor and validity of the research whatever its particular philosophy or method. This is akin to the proposal of Lee and Hubona (2009) that the fundamental logical laws of modus ponens and modus tollens can be applied across research methods to produce more rigorous research.

In the first section of the paper we develop the pluralist view of truth. Specific theories of truth are explained in Appendix A. In the next section we link truth to correctness and produce an overall framework of truth and correctness. Then in the third section of the paper we apply the framework to a variety of research approaches – positivist statistical analysis, mathematical modelling and simulation, interpretive research, critical research and finally action research. In the final section we discuss the benefits of this framework.

# PHILOSOPHICAL DEVELOPMENT OF THEORIES OF TRUTH

The issue of truth[3] revolves around two questions – does truth have a nature than can be analyzed? And, if so, what is that nature? The first question has provoked major debates between substantialist  or robust theories of truth which claim that there is an analyzable nature and deflationist theories which claim that there actually is not an underlying nature to truth, there are no mysteries to explain. The major question for substantialist theories is realism in the sense of an external world to which beliefs can correspond. In this paper we will be primarily concerned with substantialist theories since deflationist theories leave little to actually be discussed. The various theories of truth are explained in Appendix A.

[3] For good introductions to modern discussions of truth see Lynch (2001), Engel (2002) or Kunne (2003)

## Pluralist theories

Pluralist theories represent a new development in response to the stand-off between the theories described above. Generally, many philosophers do wish to maintain a substantive version of truth and do see correspondence theory as the most intuitive approach and so, in response to the criticisms of correspondence, they have developed the general idea that there may be different versions of truth dependent on the domain of knowledge concerned. There are three possible approaches (Pedersen et al. 2013a) – strong pluralism which sees only many versions of truth with no overarching unity to them, a position not held by many. Weak pluralism, which holds that truth is one and many – there is a general conception of truth, often characterized in terms of a number of properties that all forms of truth must have (called platitudes or truisms), which is realized differently in different domains. And what could be called correspondence pluralism which maintains that there is only correspondence theory but this itself can be differentially realized. Some argue that this is not properly alethic pluralism (Barnard et al. 2013).

Putnam (1994) was perhaps the first to suggest that there are many ways in which propositions can relate to reality and that therefore the word "true" may be realized differently depending on whether we are talking about physical reality, mathematics or morality. Lynch (1998) followed up with a functionalist approach asking what are the functions of truth – e.g., objective, correct to believe, and aimed at facilitating enquiry – suggesting that these functions could be met in different ways. Pedersen and Wright (2013c) provides a state of the art view of alethic pluralism (Smith 2015).

### Weak pluralism

Edwards (2011; 2013) likens truth to the notion of winning a game. We have a general idea of what winning is, but each game is different. To win at chess you need to checkmate; to win at tennis you need the majority of available sets. Thus, there is some unity of what it means to win or to be a winner that is independent of the particular game involved. And yet, determining the winner is different in each game.

In terms of truth, the unity of truth can be captured by a collection of "platitudes" such as "truth is the goal of enquiry" or "truth is a property that is distinct from justification" which describe the nature of truth in general. To see how propositions (he uses propositions as truth bearers) can come to have this truth property we then need to look at specific domains to see what it is that is accepted to generate truth in that domain. This has two aspects: first we would have to study the subject-matter of the domain to see what type it is – for example is it genuinely representational or simply discursive or logical. Second, we need to see what kind of property can establish truth in the domain, for example a correspondence between propositions and nonlinguistic entities; a coherence between linguistic entities; or a procedure or proof.

We can then form conditionals such as:

> In arithmetical discourse, if <p> coheres with basic axioms then <p> is true.

Or, alternatively:

> In arithmetical discourse, <p> is true iff <p> coheres with basic axioms.

So, a proposition that is found to be true according to the criteria of its domain is also true generally.

This approach is only a framework and there would be many details to work out:

- Can we determine an exhaustive list of platitudes to define truth in general?
- How do we decide the nature and scope of the various domains?
- Can we determine their content and criteria adequately and is there an agreed criterion for truth in each one?
- Can we show that the individual truth criteria do in fact imply the truth platitudes?

There are other versions of this approach which we will not discuss such as manifestation functionalism (Lynch 2009) and the disjunctivist view (Pedersen et al. 2013b).

### Correspondence pluralism

Within this section we will discuss two approaches – Horgan and Fumerton.

Horgan (Barnard et al. 2013; 2001) terms his approach "semantically correct assertability" (note the use of "correct" which links to the next section). Horgan is a realist accepting that there is a mind-independent and language independent world, although the world contains humans and their thoughts and activities which are clearly human-dependent. One of the things we do is make statements or assertions about the way the world is, and these statements may be right or wrong depending on how the world is, which is what we mean by truth. So for Horgan, truth is always correspondence.

However, he recognizes that, in a discourse, there are two different aspects to the way we describe or assert things about the world:

- Relevant semantics standards that govern the types of things discussed (terms) and their predicates (properties and relations) – called the "positing apparatus".

- The actual world which may or may not be as it is described.

In a small number of domains, it is possible that the terms and their predicates may *directly* correspond to elements of the real world. In these cases the semantic standards are maximally strict and we have a case of direct correspondence. However, in most discourses such a direct relation is not possible. The semantic apparatus is relatable to the world, and the world may or may not conform to it, thus truth is possible but in an indirect way. At the extreme, there may be a minimal dependence on the world and truth is defined almost entirely semantically. An example of this may be mathematics where there is only the semantic correctness with respect to mathematical axioms.

Most everyday talk, and most scientific talk, lies between these extremes with assertions about the world being semantically mediated. Consider a statement like "in 2013 Amazon was the world's biggest internet company with revenues of $88.99b" which was correct according to Wikipedia[4]. Most of the terms in this statement, such as company, revenue, even Amazon, are complex abstractions which cannot be observed directly in the world in the way that trees or tables can be. Similarly, a statement like "two is the only prime number" is true even though, ontologically, the world may not contain evenness or primeness as such.

The advantage of this approach is that under the traditional correspondence theory it was expected that there were specific truth makers holding a one-to-one correspondence with the elements of truth bearers. Under this view that is no longer necessary.

---

[4] https://en.wikipedia.org/wiki/List_of_largest_Internet_companies

*"[C]laims are true because they really do correspond to the world, even if, (as is typically the case) their positing apparatus does not map directly onto objects, properties and relations that belong to the correct ontology" (Barnard et al. 2013,p. 8)*

With this approach we can accept that much of what is said is indeed true (although discovering which statements are true is different from defining the nature of truth), and that much everyday knowledge is also true. This obviously depends on the validity of the operative semantic standards but it is likely that they will be aligned with what may be *epistemically warranted* (i.e., the pragmatic approach to truth) even though the two must not be seen as the same.

Fumerton (2013) also believes that all forms of truth are essentially correspondence between two elements, but he developed three ideas of particular interest.

First, that correspondence is not necessarily between a belief or proposition and facts about the external world. It could be between beliefs or ideas and other sets of *ideas* as, arguably, Berkeley (1995) held – thus a form of coherence correspondence. Or, it could be between beliefs or ideas and *perceptions* of the world as Hume(1967 (orig. 1750)) held, thus making the facts (truth makers) not mind-independent. Or, it could be between pragmatic utterances and the intentions and sincerity of the speaker (Habermas 1984 )[5].

Second, that correspondence truth is not all or nothing, right or wrong. There can be different degrees of truth in the relations between our beliefs and the world in the same way that pictures or models may represent with different levels of detail or faithfulness. In any case, our concepts, and even our most precise measurements always have a degree of vagueness or imprecision about them so that they cannot correspond with the world perfectly. Thus the degree of truth will depend on the form of representation (and the purpose for which it occurs). Equally, many properties, such as “tall”, are intrinsically relative not absolute. Thus Tom may be tall in general but not tall relative to the class of basketball players – these two assertions do not contradict each other, they are both true.

[5] My suggestion, not Fumerton's

Third, that there can be *different* representations of the same reality without these being necessarily incompatible, i.e., each could be true. This could occur because we investigate different aspects of the same world, e.g., through a microscope or through an x-ray, or with a painting or a photo; or it could be because we organize our observations differently perhaps because of different theoretical lenses. What correspondence theory could not accept is there being two *incompatible* pictures that are both claimed to be true: “that which cannot be stated without contradiction cannot be”.

# CORRECTNESS AND TRUTH

## Correctness

Correctness is clearly related to truth: Horgan (2001) (above) talks of truth as “semantic correctness” and Floridi (2011b) discusses a “correctness theory of truth”.  In this section we will explore the notion of correctness with a view to seeing if it might be a more appropriate term for information systems. From its definition, correctness can mean three things – true or conforming with the facts; in accordance with accepted standards; and free from error. The third is essentially the obverse of the first two so we are left with two, the first essentially as a synonym for truth, at least as correspondence, and the second wider meaning as conforming to some accepted or agreed standards or norms (Finlay 2010).

Many things may be said to be correct or incorrect: mental states such as believing (doxastic) or knowing (factive); actions or performances such as a statistical analysis or a logon procedure; representations such as a map or a computer model; information; a move in a game; or an English sentence. What is it these all have in common and how does this relate to truth?

Thomson (2008), in a major work on norms, claims that correctness has two aspects:

- Correctness is always relative to the kind of thing it is applied to. A map of England may only be correct as a map of England, not just correct in general. Correct is an attributive adjective like "good", it is always relative to a kind, K. The kind, K, fixes what properties the thing needs to have to be correct, essentially an exemplar of what the kind (or set) K would be. These properties are descriptive not normative. Note that some kinds of things do not have such exemplars, e.g., pebbles or shades of grey, and so cannot be correct or incorrect.
- At least in the case of performances or actions, there is also a normative sense of correctness – the action has to be properly performed independently of whether it succeeds or not. The golf swing may be performed correctly but the ball is affected by the wind or a bad bounce. Or, it may be performed poorly, but the ball still goes near the hole by luck.

Thompson calls the first kind external correctness (e-correctness) and the second internal correctness (i-correctness). For an action to be correct overall it must be both e- and i-correct. Sosa (2009) goes further and suggests that the e-correctness must be caused by the i-correctness – it must be "apt".

Engel (2013) considers the case of beliefs (which could of course be manifested in terms of propositions or statements) and proposes that the e-correctness is in fact truth – beliefs aim for truth and are e-correct when they are true. And the i-correctness is our evidential reasons (warrants) for believing them, which becomes normative in the sense that we *should* believe things for which we have strong evidence, whether or not they are in fact true.

This provides an interesting if somewhat circular relation to truth. From the correctness perspective, many things may be correct or incorrect but for beliefs (and their manifestation in propositions or statements) their e-correctness is a matter of truth (however construed) – if they are true they are correct. But from the truth perspective, for Horgan (2001) and Floridi (2011b),at least, truth is a matter of correctness in some form. This may appear circular, but it is a benign circularity as I shall show in the next section, in some domains truth and correctness are just the same property.

## A model for correctness and truth

This section will construct a model for combining correctness and truth based on the ideas in the above discussion.

Thomson's distinction between e- and i-correctness is fundamental and is the same as distinctions in other fields. In particular, it is essentially the same as Horgan's distinction between the semantic standards of a domain (i-correctness) and the correspondence relations between assertions and the way the world is (e-correctness). For Horgan, all (discursive) domains have i-correctness, assuming they are coherent and well-formed domains, while some have direct e-correctness and some only indirect e-correctness.

It is also the same as the distinction made within the context of model and research validation (Boudreau et al. 2001; Kleijnen 1995; Lee et al. 2003; Lee et al. 2014; Lee et al. 2015a; Lukka et al. 2010; Sargent 2013a; Venkatesh et al. 2013). Although different terms may be used, essentially there are two distinct stages to model validation which we will call verification and validation. Verification concerns the internal structure of the model, whether it is a statistical model, a simulation model, or indeed a piece of interpretive research. Validation concerns the external aspect of the model – whether it adequately represents that which it is a model of. Thus verification is i-correctness and validation is e-correctness. Within statistics and measurement theory, these are often termed precision (the degree of replicability of repeated measurements) and accuracy (the closeness of the measurement to the quantity's true value).

| Entity to which "correctness" may be applied as a type of K | i-correctness, verification, precision, normative | e-correctness, validation, accuracy, descriptive |
|---|---|---|
| Doxastic mental states, e.g., believing, guessing, hypothesizing | Whether the belief is supported by sound evidence | Whether the belief is in fact the case (truth to different degrees) |
| Factive mental states, e.g., knowing, perceiving | Not relevant except that they are coherently expressed | True by definition |
| Assertions, propositions, sentences | Whether the assertion meets the semantic standards of the domain including | Whether what is asserted is the case (truth) |
| | justification | |
| Representations, e.g., maps, pictures, models, descriptions, theories | Whether they meet the standards and norms for the type | The extent to which they correspond with that which they represent given their purpose (truth, practicality) |
| Procedures, e.g., a mathematical proof, logging into a computer account | Whether it follows to rules and axioms | Whether it succeeds (hacking an account would be e-correct but not i-correct |
| Information | Whether the signs carrying the information are semantically meaningful | Whether the content of the information is the case. This depends on what the information is about (truth of different kinds) |
| Actions, e.g., playing a game, tying a tie, performing a sonata, riding a bike | Whether they are performed in the right way according to the standards or rules | Whether they produce the right result |
| Normative kinds of artefacts which may be good or defective, e.g., machines, computer systems | Whether it exemplified the standards appropriate to its kind (form) | Whether the object has the properties that would make it a "good" example of its kind (function) |

Table 1 The concept of correctness applied to different domains

We will briefly explain the model here and then look at particular parts in more detail in the context of information systems. In Table 1, the first column give a (non-exhaustive) list of entities or types to which the term correctness could be applied (correctness-bearers!). The next two columns describe the relevant forms of i- and e- correctness.

Representation cover a range of things that may have different purposes – a picture or photo may just describe something; a theory may explain why something happened; a simulation model may try and replicate behavour. Whether they are e-correct depends both on their correspondence but also the purpose – a map of the London underground is not correct in terms of walking the streets.

Procedures are specified steps that need to be undertaken to achieve a result. A mathematical proof does not correspond to anything but, starting form axioms and following logical rules it can generate a conclusion. It is i-correct in terms of adhering to the rules and potentially other criteria such as elegance or simplicity. It is e-correct in demonstrating the result. Logging into an account is e-correct if it succeeds. It may not be i-correct if done in the wrong way, for example by hacking.

Information is a disputed phenomenon with different conceptions (McKinney et al. 2010). Some, such as Floridi (2011a) and Mingers and Standing (2014a), argue that information is both objective and true in which case its i-correctness is its semantic meaningfulness and its e-correctness is its truth in some form. Others (Checkland et al. 1998b) argue that it is subjective and not necessarily true in which case it is difficult to understand what might be correct information.

There are many artefacts, especially humanly produced, which may have the property of "being a good K" if there are standards or properties which exemplars of such a kind exhibit. A good toaster produces evenly browned toast; a good information system produces accurate, timely, relevant information in an easy to use way. For these, the e-correctness involves meeting the specified goodness criteria, which may often be in terms of functions. The i-correctness concerns the form of the artefact – is it aesthetically pleasing? Is it robust and easy to use? In these cases there could be disagreements about which properties were part of the function and which the form.

# APPLYING THE FRAMEWORK TO INFORMATION SYSTEMS

We began by highlighting the importance of knowledge for IS, and arguing that knowledge is truth-constituted, i.e., it must be true to be knowledge. This led us to the concept of truth and we investigated various theories of truth, especially the pluralist versions of truth which see it and many and one. From here, we considered the intimate connections between truth and the wider concept of correctness, which could be said to subsume truth. We developed a framework for analyzing the nature of internal and external correctness in several domains. In the rest of the paper, we will apply this framework to a range of concepts and phenomena within information systems.

In actuality, the concept of truth itself is seldom discussed in IS research papers (Becker et al. 2007; Webb 2004), although there are many papers that debate the nature of validity for different forms of research (Johnson et al. 2006). As we said earlier, there are two distinct questions concerning truth – what is it, i.e., what is its nature? And how do we discover it, i.e., how do we tell true theories from false ones? We may call these the definitional and the justificational questions. The first part of the paper has been concerned with the first question, but we now move to the second, practical question – how do we justify our theories?

## Justifying quantitative research

Within IS, there is a considerable literature devoted to justifying empirical, quantitative research. Indeed there is strong line of good practice recommendations developing, in the main, from Cook and Campbell's (1979) treatise on quasi-experimentation. This was picked up by Straub (1989) and further developed in theoretical (Bagozzi 2011; Im et al. 2015; MacKenzie et al. 2011; Shadish et al. 2002; Straub et al. 2004; Venkatesh et al. 2013) and empirical studies (Boudreau et al. 2001; King et al. 2005). Theoretically, this approach ties in directly with our correctness framework:

"We use the term *validity* to refer to the approximate truth of an inference. When we say something is valid, we make a judgement about the extent to which relevant evidence supports that inference as being true or correct … Validity is a property of inferences. It is *not* a property of designs or methods" (Shadish et al. 2002, p. 34)

Whether it does in practice is more debatable.

The original work (Campbell 1957) distinguished between internal and external validity (cf internal and external correctness above) while Cook and Campbell (1979) added construct validity and statistical validity. Straub et al (2004) develop these as shown in Table 2 (construct validity becomes a part of instrument validity).

| Validity type | Meaning | Means of assessment |
|---|---|---|
| Instrument validity | Assesses the validity of the research instrument, typically a questionnaire or experiment | |
| • Content | Do the instrument | Literature review, |
| | measures adequately reflect the content of the construct they are measuring? | expert judgement |
| • Construct<br>○ Discriminant<br>○ Convergent<br>○ Factorial<br>○ Nomological<br>○ Predictive | Do the measures converge on the construct and not on other distinct constructs? | Statistical methods such as CFA, SEM, PCA;<br><br>Judgmental comparison; Quantitative comparison |
| • Reliability<br>○ Consistency<br>○ Test/Retest<br>○ Split half<br>○ Inter-rater | Are the results/responses repeatable? | Cronbach's alpha |
| Internal validity | Are there alternative causal explanations for the observed data? | Not discussed |
| Statistical validity | Are the results sufficiently statistical robust that they are unlikely to have occurred by chance? | Rsquared, F, SEM<br>See (Gefen et al. 2000) |
| External validity | To what extent can the findings be generalized to other populations and settings? | Not discussed |

Table 2 Four forms of validity for positivist research summarized from Straub et al (2004)

These forms of validity take on quite specific meanings. First, we should note that although the title of the paper (Straub et al. 2004) is “Validation guidelines for IS positivist research”, which is quite general, in fact the guidelines only refer to specific forms of statistical research in which there are some underlying latent, subjective constructs, and relationships between them, which are then operationalized in terms of particular quantitative measures and an instrument to collect data. The instrument is assumed to be some form of questionnaire or perhaps experiment. The title of Straub’s (1989) earlier paper, “Validating instruments in MIS research” is perhaps more accurate. The point is that there are other forms of quantitative research beyond surveys and statistics, popular as they may be.

Moreover, most of the discussion concerns the fairly technical issues of instrument validity and statistical validity rather than the more general ones of internal and external validity. Again, these latter concepts are defined quite narrowly, and perhaps counter-intuitively, in this approach – internal validity only concerns the possibility of there being other causal relationships, i.e., explanations, that are not included in the model. In many ways this seems like an external factor since it makes direct reference to the external worlds beyond the model, and cannot really be dealt with from a purely internal perspective.

Equally, the idea that external validity primary concerns the extent to which the results can be generalized to other populations and settings (King et al. 2005) seems mistaken. As Reichardt (2011) argues, the fundamental purpose of validation is to assess the truth of the inferences made in the model, it is not particularly concerned with how wide or narrow those inferences are.: “As long as a generalization about a causal relationship is true, it is externally valid even if the generalization is exceedingly narrow” (p. 46).Whilst generalizability is an important and much debated (Lee et al. 2003; Lee et al. 2012; Seddon et al. 2015; Tsang et al. 2012), characteristic of a statistical finding, it is a separate issue from the question of validity.

We should also note that this approach to validity does not properly separate validity from precision (Reichardt 2011). One of the fundamental distinctions in statistical inference is that between accuracy and precision. An estimate or inference may be accurate but imprecise (having wide confidence intervals) or it may be inaccurate but precise. Validity concerns the accuracy of the inference rather than its precision but these are conflated in the validity typology.

Finally, this approach makes almost no reference to the fundamental issue of designing the study in the first place in such a way that the eventual results will form valid answers to the research questions. It takes for granted the development of appropriate constructs, hypotheses of the relationships between them, and the initial determination of the appropriate measures and data collection instrument and yet arguably these factors are much more important for overall validity or correctness of the research findings that is instrument validity (Johnston et al. 2010). As the empirical research shows (Boudreau et al. 2001; Jones 2004; King et al. 2005; Straub 1989), in many cases of papers published in leading journals even the most basic aspects such as describing and justifying the methods of data collection and analysis are absent.

Lee and Hubona (2009) provide an alternative approach to validation. Their primary aim is to produce a framework that can apply to both qualitative and quantitative research based on the logical forms of argument – modus ponens (p implies q; p; therefore q) and modus tollens (p implies q; not q; therefore not p) which they call the MPMT framework. They distinguish between formative validity and summative validity (taking these terms from education research) and suggest that much IS research involves formative validity but little summative validity. Formative validity is the process of forming or producing the theory or inference and so this type of validity concerns the extent to which the research has correctly followed an accepted procedure. Summative validity is a characteristic of the sum result or product of the process that has been followed. It involves comparing the consequences or predictions of the theory with observed evidence according to the logic of modus tollens. If a consequence or prediction of the theory cannot in fact be observed then the theory does not have summative validity and could potentially be rejected. Lee and Hubona show that this approach can apply to quantitative research, qualitative research, and even systems design – a system may be designed according to an accepted systems design methodology and yet still fail to meet its aims. They also argue that of the two, summative validity is more important than formative validity even though in practice it is seldom demonstrated, particularly in positivist research.

In order to generate summative validity in statistical-type research (which is the content of this particular section), Lee and Hubona argue that statistical validity in the sense of significance tests or confidence intervals for various fitted parameters which constitute the hypothesized relationship is not sufficient. This is actually part of formative validity. As well as this it is necessary to test the theory's predictive capabilities on out-of-sample data points using hold-out samples or cross-validation. We should note, however, a very common problem pointed out by Lee and Hubona – the fallacy of affirming the consequent. If we find that the predictions are in fact correct, does that prove or confirm the theory? The answer is unfortunately no, since there could always be some other explanation which actually accounts for the results. This can be expressed in logic – p implies q; q; therefore p – which is not a valid inference. This point relates to Straub's issue of internal validity which concerns alternative explanations. We would suggest that this is mis-named and is really external or summative validity – as well as trying to confirm the predictions one also need to actively try and eliminate alternative explanations (cf. the section on critical realism below).

In comparison with the correctness framework, it seems clear that Lee and Hubona's approach fits it very well. Formative validity is essentially the same as internal correctness, while summative validity is the same as external correctness and the two are related but independent. We would hope that formative validity (i-correctness) would lead to summative validity (e-correctness) but it is not guaranteed; while it would be possible to reach summatively valid conclusions even through research that was formatively weak.

**Other forms of quantitative research**

The previous section was primarily concerned only with statistical type research but there are many other forms of quantitative research of potential relevance, for example simulation or mathematical modelling. In this section we will briefly consider simulation as representative of these.

Simulation involves building a computer model that is intended to replicate the behavior of a real-world system of interest. There are three major types – discrete event (DES), system dynamics (SD) and agent-based modelling (ABM) – which employ different modelling techniques but are similar in terms of validation. Simulations are generally developed for a specific purpose – better understanding of a system, improvement of the system's operations, or the design of a new system – and therefore involve decision-makers and others affected by the results. It is important for these stakeholders that they have confidence in the correctness of the simulation and its results (Sargent 2013b). Although there is a degree of debate, the correctness of a simulation is generally evaluated in terms of verification and validation (Robinson 1997) although its credibility with users is also important (Robinson 2002).

Sargent (2013b) illustrates these concepts in terms of three elements – the object system that is to be simulated, the conceptual model of that system, and the computerized version of that conceptual model. Verification then concerns the correctness of the model and its computer implementation while validation has several components – conceptual model validation that the conceptual model is a correct representation of the object system; operational validation that the outputs of the computer model are sufficiently accurate with respect to the object system for the purpose at hand; and data validation that the available data is sufficiently correct for model building, evaluation and testing.

The three forms of validity are independent but inter-related. If the conceptual model is invalid, than it is unlikely that the final model will have operational validity. If appropriate valid data is not available then a valid conceptual model could be built but then not operationalized. It is also important to emphasizes that validity is not absolute but always relative to the purposes of the simulation exercise – to understand puzzling behavior a fairly simple model may be sufficient, but to help operate a complex production plant the model may need to be highly detailed and complex[6]. The general advice is to keep the model as simple as is possible to meet the objectives (Robinson 2007). Credibility depends to some extent on validity – a verified and validated model should generate credibility – but special steps may be taken to improve it, for example participation by stakeholders in the development process and techniques such as hi-res animated graphical outputs.

[6] The point that validity is not absolute but relative to purpose also applies to the statistical modelling discussed above but is seldom mentioned.

There are many techniques and tests used in verification and validation (Sargent 2013b), for example for verification there are comparison with other models, extreme conditions tests, degeneracy tests, sensitivity analysis, replications, trace tests, while for validity there are predictive validation, comparison with historical data, event validation, face validation, graphical animation or structured walkthrough.

In terms of our framework then clearly verification is i-correctness and the various forms of validation are e-correctness. Credibility is interesting in that it could be regarded as separate from the correctness of the model, or it could be regarded as the ultimate form of e-correctness – if the model is not believable for the clients then it fails no matter how good it was. These forms of validity can also be related to the different theories of truth (Becker et al. 2005; Schmid 2005). Clearly the primary forms of validity rely on correspondence theory; verification, especiually comparison with other models, can be seen as coherence theory, and the issue of credibility can be seen as pragmatic – rationally/consensus – rationally acceptable under ideal epistemic conditions.

## Justifying qualitative research

Qualitative or interpretive research is a much more complex area in terms of validation and truth (Cole et al. 2007; Goldkuhl 2012; Myers et al. 2002a; Myers et al. 2002b; Walsham 2006b). First, there are a wide variety of methods that differ significantly in their ontological and epistemological assumptions from relatively objective post-positivist approaches such as grounded theory (Glaser et al. 1967) through textual analyses such as semiotics (Mingers et al. 2014b; Mingers et al. 2014c) or discourse analysis (Cukier et al. 2009)to highly subjectivist ones such as phenomenology (Mingers 2001b). Second, there is debate even within methods as to the possibilities of some form of external validation and some researchers would deny this possibility altogether. Papers in IS that provide guidance on doing interpretive research generally fail to discuss validity. For example, Klein and Myers' (1999) authoritative paper provides seven principles that should be applied in interpretive research (primarily limited to hermeneutics) but say little about validation principles. Similarly, Sarker et al (2013) review empirical studies and also offer guiding principles but do not discuss validation.

Interpretive research begins from the position that its object of study, whether it is actions, texts, beliefs or discourse, is socially constructed by the actors involved. Therefore, its primary task is to gain an authentic understanding (*verstehen*) of that meaning in the terms of the actors who produce it rather than in terms of theory, or the interpretations of the researchers. For some researchers, e.g., ethnographers, that is sufficient whereas others would want to go on and interpret the results and perhaps relate them to theory.

Moving to possible validity criteria, some of the first were proposed by Lincoln and Guba (1985; Shenton 2004) as a direct analog to the criteria for positivist research discussed (and criticized) above: internal validity – credibility; external validity – transferability; statistical validity – confirmability; and reliability – dependability. They later argued that (Lincoln et al. 1986) these criteria are overly influenced by the concerns of positivist research and suggested that these four constituted the *trustworthiness* of research but other conditions concerned with the wider application and results of the enquiry were also needed which they called termed *authenticity*.

Maxwell (1992) suggested three forms of criteria based partly on different stages of the project. First is descriptive validity which solely concerns the quality of the data production process – that it is comprehensive, accurate and not subject to dispute (although the participants may themselves hold different and perhaps contradictory viewpoints, these should be faithfully recorded). The second is interpretive validity which goes beyond merely recording events, actions and discourse to generating interpretations of it, but still from the participants' point of view not the researchers. This has been described as an "emic" viewpoint rather than an "etic" viewpoint (Headland et al. 1990), an insider rather than an outsider one. Interpretive validity involves the faithfulness or authenticity of the account to those involved, but even here the boundaries are blurred because actors are not always fully transparent to themselves and, as Giddens (1979) emphasizes, there are often unknown conditions and motivations for action. The third form of validity is theoretical validity which does move away from an emic account to an etic one. The researcher aims to develop *theories* that may explain the particular observed behaviors. Theory could come in two directions, from within as in the case of grounded theory where the theory is developed internally from the research material, or from without as theory that already exists is applied to explain the situation. [7]

Lee and Hubona (2009) argue that their MPMT framework applies equally to interpretive research. They give the example of the hermeneutic understanding of a text whereby (in terms of summative validity) if the researcher has a correct interpretation of the text then it should be consistent with any particular passage or set of passages (MP). But, if a contradiction arises then that implies that the interpretation is not correct (MT). They suggest that this approach is a realization of the hermeneutic circle. This is the only actual example they give, but they do analyze a set of interpretive papers and find that all but one only discuss formative validity, in terms of the processes employed, and do not try and test their interpretation in a summative way. For testing summative validity in research approaches other than hermeneutics they follow Sanday's (1979) and Schutz's (1962) proposal that it should be understandable and acceptable to actors in the situation and potentially enable a stranger to act appropriately within the culture. This can be termed authenticity.

[7] Maxwell does discuss two other forms of validity – generalizability and evaluative validity. We consider the former, as argued in the section on quantitative research, to be orthogonal to the primary question of truth and validity. The question of evaluation, i.e., judging actions to be right or wrong, will be considered further in the section on critical research.

Venkatesh et al (2013), based on a consideration of several of the above typologies, suggest another three fold classification: i) design validity (which includes descriptive validity, credibility and transferability); ii) analytical validity (including theoretical validity, dependability, consistency and plausibility); and iii) inferential validity (including interpretive validity and confirmability). In comparison with Maxwell's typology, this one seems rather confusing to us. Design validity actually includes elements descriptive and interpretive validity (i.e., concerned with the validity of the process) mixed with generalizability. Analytical validity seems to include elements of both interpretive validity and theoretical validity, while inferential validity seems to go back to interpretive validity rather than to inferences beyond the situation.

This whole area is clearly complex and confused in its terminology. From our validity and correctness point of view we wish to have a classification which is quite general and compatible with many of the particular approaches. We would therefore make one main distinction, that between emic and etic research. In emic research (which must necessarily come before etic) the primary concern is with reproducing, in as *authentic* and rich a manner as possible, the way of life of the actors within a situation of interest, in their own terms. This includes both descriptive and interpretive validity in Maxwell's model. Some research, for instance ethnography, may choose to stop there but increasingly there is a view that even ethnographic research should move towards some form of explanation (Kakkuri-Knuuttila et al. 2008; Lukka et al. 2010). This etic account, either based on or generating theory, will be expressed in the researchers' language and must be *plausible* to the research community. This is congruent with Cole and Avison's (2007) use of the trustworthiness of the research process and the truthfulness of the results. Based on our correctness framework, both of these aspects of research will have both i-correctness (formative) and e-correctness (summative) validity criteria as shown in Table 4.

## Justifying system design: design science and action research

We are considering these two somewhat different approaches together for two reasons. First, they share purposes that make them different from the research approaches we have so far considered – that is, they both aim to bring about beneficial change in organizations, one through the development of an IT artifact, the other through general activity which might include developing artifacts. Second, because these similarities have already been noted in the literature (Baskerville et al. 2009; Järvinen 2007; Lee 2007; Sein et al. 2011; Wieringa et al. 2012) although Iivari and Venables (2009) suggest the similarities may not be deep. But from the point of view of validation they do have commonalities.

Design science is concerned with producing new and innovative IT artifacts to solve organizational problems (Hevner et al. 2004) although Lee et al (2015b) point out that it should be the IS artifact not just the IT artifact. As Hevner et al ((2004, p. 78) note, "design is both a process (set of activities) and a product (artifact) – a verb and a noun" and this concords with the two aspects of correctness in our framework – i-correctness (formative) as conforming to a process or methodology, and e-correctness (summative) as successfully achieving its goal or purpose (in this case in terms of its organizational stakeholders).

Various proposals have been made for a design science methodology that has been integrated by Peffers et al (2007) into the following:

1. Problem identification and motivation
2. Define objectives for solution
3. Design and development
4. Demonstration
5. Evaluation

Note that here the 5th step is actually evaluation, in particular evaluating whether the artifact does indeed meet the objectives that were required of it (summative or e-correctness).

Venable et al (2012) have developed a detailed framework of different methods for assessing both formative and summative validity. It is based initially on the 5E's approach to evaluation (Checkland et al. 1990) – Efficacy, Efficiency, Effectiveness, Elegance and Ethicality. Of these, Efficacy and Effectiveness primarily concern summative validity and the other three concern formative validity. Efficacy is the extent to which the artifact performs as it is designed to do whilst effectiveness is the extent to which performing those tasks is actually successful in the organizational context - does it do what it is supposed to do, and is that the right thing to do?

In terms of formative validity, first was the artifact designed according to a rigorous methodology of whatever kind? Then there are questions as to whether it was developed with an economical use of resources (efficiency), according to ethical principles, and ultimately elegantly and aesthetically (the Mac vs the PC?)? One would like to think that formative validity would lead to summative validity but unfortunately the high number of IS failures that still happen (Dwivedi et al. 2014; Georgiadou et al. 2006) shows that this is not the case. These forms of validity are shown in Table 4.

Moving to action research, although as we have shown there are many who consider the two can be intimately linked, in terms of validation we will deal with them separately although Wieringa and Morah (2012) actually define the concept of "technical action research" as a specific method for evaluating design science. Action research (AR) (Checkland et al. 1998a; Eden et al. 1996) has a long history dating back to Kurt Lewin (1946) and comes in many varieties including action learning (Revons 1993), action science (Argyris et al. 1985) and participatory action research (Whyte 1991). It has been recommended for research in information systems (Baskerville et al. 1998; Baskerville 1999; Chiasson et al. 2009; Davison et al. 2004).

Given this variety we will have to consider a very broad description of AR as being constituted by several elements performed in a cyclical manner:

1. Initial recognition of problematic issue and entry of researcher
2. Declaration of theories and methodologies thought to be relevant
3. Undertaking action to improve the situation as both participant and researcher (in participatory AR the actors are also seen as participant researchers)
4. Evaluate results in terms of improvement to the particular organizational situation

5. Evaluate results in terms of the theory/methodology used and disseminate

In terms of the correctness or validation of the process (Baskerville et al. 1998; Checkland et al. 1998a; Eden et al. 1996), we can consider the i-correctness in terms of the extent to which the AR process was followed, and the e-correctness in terms of two distinct criteria – the success in terms of resolving the problem, and the learning and development of theory which may be applicable elsewhere. It is the latter which mainly distinguishes action research from pure consultancy. Checkland and Holwell (1998a) emphasize the importance of "recoverability", that is explicit documentation of the process followed and decisions made which will help generate the theoretical lessons as well as allowing later critical scrutiny.

## Justifying a critical approach

In this section we will cover a range of explicitly critical approaches mainly based on the work of theorists such as Bourdieu (Kvasny et al. 2006), Foucault (Willcocks 2004) and Habermas (Brocklesby et al. 1996; Howcroft et al. 2005; Klein et al. 2004; Myers et al. 2011); as well as critical realism (Johnston et al. 2010; Mingers 2004; Mingers et al. 2013), and critical versions of interpretive approaches such as critical ethnography (Myers 1997) and critical discourse analysis (Cukier et al. 2009). A critical approach, or the idea of critique, has two lineages one traceable to Kant and one to Marx (Cecez-Kecmanovic 2011; Mingers 2000). Kantian critique concerns the limits of our knowledge and research methods while Marxist critique concerns the oppressive nature of society. Generally, both are involved in a critical approach. However, a critical approach is not primarily about research *methods* but about attitude and values (Cecez-Kecmanovic 2011; Morrow et al. 1994). In other words, there are not specific critical research methods, rather traditional methods, both quantitative and qualitative, are used but with a critical intent.

Alvesson and Deetz (2000) provide perhaps the most general framework for doing critical research that involves three stages[8] :

- Insight – hermeneutic understanding and the archaeology of knowledge. This stage involves gaining knowledge and appreciation of the situation of interest using a range of ordinary research methods, both qualitative and quantitative. But it will be guided by explicitly critical attitudes and values and will view the subjects as active participants in the research rather than passive objects.
- Critique – deconstruction and the genealogy of knowledge This stage involves using varied critical theories and constructs to uncover and reveal the often hidden or suppressed mechanisms that distort the participants understandings of the situation and act so as to maintain this power differential.
- Transformation redefinition – enlightenment and emancipation. This stage aims at enlightening participants to the true nature of the situation and thereby helping them to bring about change. It also reflexively develops social theory. Final validity is in the judgement of the participants.

Table 3 shows a range of critical research approaches and how they can be mapped to Alvesson and Deetz's (2000) three stages.

---

[8] We have developed around their actual criteria

| Alvesson and Deetz (2000) | Myers and Klein (2011) | Cecez-Kecmanovic (2011) | Johnson et al (2006) based on Kincheloe and Mclaren (2005) | **Critical realism (Bhaskar 1994; Mingers 2009; Mingers 2014)** |
|---|---|---|---|---|
| Insight | Interpretive research | Critical understanding<br>• Critical theory concepts<br>• Emancipatory values<br>• Choice of research methods | Research designs that are participative and democratic, and approximate Habermas's ideal speech situation.<br>Reflexive analysis of researchers' interests and assumptions | Science is value-laden not value-free and should be used to understand the true nature of society |
| Critique | Utilize critical theories<br>Explicitly adopt social values<br>Reveal and challenge the status quo | Critical explanation and generalization<br>• Hidden mechanisms<br>• Wider contextualization<br>• Social and power relations | Critical ethnography to sensitize researchers and participants to how society distorts the subjectivities of participants<br>Comparison of particular context with other comparable ones | Explanatory critique - a critique of the false beliefs held by social actors and the social/ organizational structures that maintain them |
| Transformative redefinition | Emancipation<br>Improve society<br>Improve social | Open discourse<br>• Non-distorted communication | Catalytic validity – the extent to which the research | Theory practice consistency – given the |
| | theory | • Transformative praxis<br>Reflexive dialectic | changes participants self-understandings and thereby enable them to change the situation<br>Credibility for participants is vital | explanatory critique, this should lead to action dedicated to removing the constraints and ills<br>Then universalizing this to similar constraints and problems in other contexts |

Table 3 Different approaches to critical research

From this we can see that e-correctness concerns the actual success of the critical analysis in terms of the change of consciousness of the participants, and change of oppressive social arrangements. This is ultimately to be judged by the participants themselves rather than the researchers. I-correctness concerns the process of research and analysis itself and whether it has properly followed the research steps as described in Table 3. This is summarized in Table 4.

| Form of research | i-correctness, normative, verification, precision, formative | e-correctness, descriptive, validation, accuracy, summative | Relevant forms of truth |
|---|---|---|---|
| **Behavioral statistical research** (positivist as | Formative validation:<br>• Content<br>• Construct | Summative validation:<br>• Comparison of predictions with actuality, e.g., | **Correspondence** between constructs and concepts, and between results and |
| defined by Straub (1989) and Lee and Hubona (2009)) | • Reliability<br>• Statistical validity | cross-validation<br>• Elimination of alternative explanations | actuality;<br>**Coherence** of constructs |
| Simulation and other mathematical modelling | Verification:<br>• Model comparison<br>• Extreme conditions<br>• Degeneracy tests<br>• Sensitivity analysis<br>• Replications | Conceptual and operational validation:<br>• Predictive validation<br>• Historical data<br>• Event validation<br>• Face validation<br>• Graphical animation<br>Credibility | **Correspondence** between results and actuality;<br>**Coherence** of model;<br>**Pragmatism** and **consensus** about operational validity |
| Interpretive research<br>-<br>- emic | Formative validation:<br>• Descriptive validity<br>• Interpretive validity<br>• Consistency<br>• Credibility<br>• Dependability | Summative validation<br>• Authenticity in the eyes of the participants/subjects<br>• Performativity – the stranger test | **Correspondence** between description and participants world;<br>**Consensus** about authenticity of results |
| - **etic** | • Theoretical validity | • Plausibility in the eyes of the research community | **Consensus** about plausibility of |
| | | | theoretical interpretation |
| Design research | • Methodological validity<br>• Efficiency<br>• Ethicality<br>• Elegance | • Efficacy that the system works<br>• Effectiveness that it does the right thing | **Pragmatism and consensus** about operational success<br>**Coherence** of design method |
| Action research | • Declaration of theory and methodology<br>• Active application of theory and participation in situation<br>• Recoverability | • Effectiveness that the problematic issue has been alleviated<br>• Justification of theoretical contribution<br>• Generalizability to other contexts | **Pragmatism and consensus** about operational success<br>**Consensus** about plausibility of theoretical learning<br>**Coherence** of results with methods used |
| Critical research | • Critical perspective and use of critical theories<br>• Participative research design<br>• Analysis of underlying, coercive mechanisms<br>• Comparison with other context<br>• Researcher reflexivity | • Enlightenment of individual participant<br>• Change of social arrangements<br>• Judged by the participants | **Correspondence** of theory to social mechanisms<br>**Pragmatism and consensus** about enlightenment and change |

Table 4 The correctness framework of IS research

# DISCUSSION AND CONCLUSIONS

To begin with we would like to make it clear what we are *not* suggesting. First that we are not prescribing or privileging any particular research methods, indeed our whole argument is based on the idea that all these approaches (and others we have not covered) may well be able to contribute to information systems whether it is in terms of knowledge, understanding, or practical developments. This is true for a single research method or for a combination of methods within a multimethodology (Mingers 2001a; Mingers 2011; Venkatesh et al. 2013), which is actually our preferred option. Second, we are not suggesting direct changes to specific research methods, whether it is the statistical analysis of surveys or the coding of ethnographic data, but we are suggesting that they should be carried out with more concern for, and consideration of, their validation.

The implications of our analysis of truth and correctness are as follows. Research is often carried out and published with little explicit regard for its validation (Boudreau et al. 2001; Jones 2004; King et al. 2005; Straub 1989) . The main argument of this paper is that this is not acceptable. For research to make a genuine contribution, either to knowledge or to practice, and to be published in journals or lead to organizational change, every effort must be made to demonstrate that the results are valid, that is believed to be *true* or *correct*.

As we have demonstrated, there are two fundamental and distinct characteristics – internal correctness and external correctness, also known as verification and validation or formative and summative. The first is normative and concerns the way in which the research has been carried out; the second is descriptive and concerns the relationship of the research findings to the external world. We have shown within the framework criteria for both of these across a wide range of research approaches. We agree with Lee and Hubona (2009) that much less attention is paid to e-correctness than i-correctness and yet arguably the latter is more important.

It is tempting also to align these two with rigor and relevance. Certainly i-correctness concerns the rigor of the research, and e-correctness is at least related to its relevance although there may be very abstract research which does not, at the time, seem to have much direct relevance but one only needs to think of the laser or prime number theory to see how such research may later come to have huge relevance.

The way in which research does need to change is that it needs to explicitly consider both these aspects of correctness at all stages – the design of the research, its operationalization, and its description and dissemination. We hope that the framework can provide a checklist for researchers to consider in designing their research, and for referees and editors to look for when evaluating submissions or grant applications.

We believe that it is important that we have produced a framework that encompasses a wide range of methods. Too often, different research methods are seen to be in competition or even in conflict with each other. The framework demonstrates that they can all be seen as sharing some very basic characteristics, and are all ultimately part of the same human drive to better understand and improve the world. By focusing explicitly on both internal and external correctness we hope that the results of research will be more informative and effective.

In terms of limitations and further research, we note the following. The framework could be developed to include further research approaches that we have not considered, for example theoretical computer science or feminist research or multimethodology. It could also be developed internally to provide a greater discrimination within approaches, especially the interpretive area where it may be found useful to have different criteria for, say, hermeneutics, phenomenology, textual analysis or semiotics. The advantage of a framework such as this is that it makes everything explicit (Klein et al. 1999) so that it can act as a trigger for debate. It may well be that proponents of particular methods may disagree with our validity criteria but at least there is now a target to be aimed at.

For a major research question it may be that all the validity criteria cannot be answered within a single study – there may need to be sequential studies, perhaps some formative, and then later ones summative; or different methods may need to be applied to different aspects of the situation, thus invoking different validity questions. These considerations clearly lead on to the possibility of mixed methods work. They also touch on the question of generalizability. In this paper we have distinguished the e-correctness (or validity) of a particular study from the extent to which it can be generalized to other contexst, but the two are clearly related, and the generalization question raises its own validity issues that we have not here addressed.

# APPENDIX A TRADITIONAL THEORIES OF TRUTH

## Robust truth theories

### Correspondence Theory

The essence of correspondence theory is what is called "alethic[9] realism", that is that truth depends on the way the world actually is, so truth has a nature and its nature is objective – it depends on the world itself, not what we believe about it. Correspondence theories involve specifying what may be true (truth bearer), the "reality" to which it corresponds (truth maker), and the nature of the correspondence relation. There have been a variety of answers to these questions as shown in Table 1.

| Theory | Truth bearer | Relation | Truth maker |
|---|---|---|---|
| Russell (1906) | Beliefs | A structural isomorphism between the belief and the facts | Facts – a complex unity of parts and relations |
| Austin (1950) | Propositions or sentences | Correlation, sometimes conventional, rather than structural isomorphism | Things, features, facts, states of affairs |
| Field (1974) | Words or sentences | A causal relation – states of affairs lead us to make particular statements | The world |
| Alston (2001) | Propositions – the content of | Objective, mind-independent, non- | Facts about how the world is |

[9] From the Greek meaning related to truth

| | | | |
|---|---|---|---|
| | the act of stating or believing | epistemic (i.e., not based on our evidence) | |

Table 1 Examples of correspondence theories

These different versions of correspondence theory all share the core concept that there is something, the mind-independent world, that makes our beliefs or propositions true whether or not we can discover or justify that truth.

There are many objections to correspondence theory. In brief:

- The consistency problem -that beliefs or statements are different kinds of thing to states of affairs or facts in the world and that the two cannot logically be compared. Beliefs can only be compared with other beliefs.
- The realism problem - that we do not have epistemological access to an independent external reality, we always experience it through our perceptions, cognitions and language, and so we could never discover if our beliefs are true
- The justification problem - that the truth of a proposition is independent of our justification for it so all our beliefs could be false.
- The scope problem - that propositions could be of so many different kinds (scientific, mathematical, fictional, moral etc.) that there can be no one property or causal relation than makes them all true

These arguments have led to the main, substantive, alternatives to correspondence.

**Coherence theory**

Coherence theories differ in terms of both the truth relation and the form of truth maker. In general, coherence theories specify that the relationship is one of internal consistency and coherence with some set of other consistent propositions or beliefs rather than any reference to an external world. Theoretical holism (Quine et al. 1978) requires that a belief is logically consistent with some system of beliefs but does not specify precisely what that set might be. Joachim (1906) argued that that the set of beliefs must form a comprehensive and significant systems of beliefs, and Blanshard (1941) went further saying that such a system should be comprehensive in including all known facts, and where each judgement should entail and be entailed by every other. More recently, Alcoff has suggested that the system should not consist just of beliefs but also of social practices, traditions and life events. And, moreover, that there can be different sets of beliefs, accounting for different experiences of the world, which may not necessarily be contradictory.  Coherence theory has been applied in specific domains such as mathematics.

**Pragmatist and consensus theories**

These theories judge truth not in terms of correspondence to reality, but in terms of the degree of evidence, agreement or usefulness. For this reason they are called epistemic theories. They can be traced to the American pragmatist philosophers. For instance, truth for Peirce (1878) was "the opinion which is fated to be ultimately agreed to by all who investigate".  For James (1976) what is important is what practical effect truth would have - "true ideas are those that we can assimilate, validate, corroborate, and verify. False ideas are those that we cannot".  Dewey (1938) introduced the idea of "warranted assertibility": "If inquiry begins in doubt, it terminates in the institution of conditions which remove need for doubt. The latter state of affairs may be designated by the words belief and knowledge. For reasons that I shall state later I prefer the words 'warranted assertibility'". Thus, from this perspective, there is not an absolute truth, certainly not in correspondence with an external reality. Rather, truth (or perhaps knowledge) is always provisional and fallible, based on the best evidence and information that we have, and moving towards but never perhaps reaching, the ideal of certainty.

In more recent times, Putnam (1981) was close to this view in arguing that truth was what we would agree on under ideal epistemic conditions, "ideal warranted assertibility" which would depend on the particular entities being studied and was an ideal in the sense that it could be approached but never realized in practice (he later moved away from this approach). Wright (2009) has proposed the alternative notion of "superassertibility". For Peirce and Putman, getting closer to the truth involves gaining more and more precise information under increasingly ideal conditions. Wright suggests instead that, given some reasonable and practical evidence or information in favor of an idea, we should ask, would it remain warranted no matter how the information was improved or enlarged in the future.

"A statement is superassertible, then, if and only if it is or can be warranted and some warrant for it would survive arbitrarily close scrutiny of its pedigree and arbitrarily extensive increments to, or other forms of improvement of, our information" (Wright 2001, p. 771).

This notion has been suggested as applicable to domains such as ethics.

Habermas, too, has had an essentially pragmatist/consensus theory of truth but he has changed in a significant way more recently. Originally, with his theory of knowledge constitutive interests (Habermas 1978), he identified three forms of science – empirical/analytic, hermeneutic (normative) and emancipatory – but all three were underpinned by a discursive theory of truth. Like Putnam, he discussed the circumstances under which ideal agreement could be reached and called the “ideal speech situation” where the truth would be generated by “the unforced force of the better argument” [Habermas, 1974 #1516, p. 240; Habermas, 2003 #2094, p. 37]. So at that point, truth was identified with that which would emerge through infinite, unfettered debate.

However, he now [Habermas, 2003 #2094] recognises a substantive difference between the empirical domain and the normative domain. Whereas normative or moral issues can only ever be established through debate and discourse, propositional statement about the material world can now be proved wrong by events even if they were to be the result of an ideal debate.

“I have given up an epistemic conception of truth and have sought to distinguish more clearly between the truth of a proposition and its rational assertability (even under approximately ideal conditions).” [Habermas, 2003 #2094, p. 8]

Habermas now accepts the basic realist view that there is a world independent of human beings; that we all experience the same world; and that this places constraints upon us;, whilst still accepting that our access to this world is inevitable conditioned or filtered through our concepts and language.

“These objections have prompted me to revise the discursive conception of rational acceptability by *relating* it to a pragmatically conceived, nonepistemic concept of truth, but without thereby assimilating ‘truth’ to ‘ideal assertability’” [Habermas, 2003 #2094p. 38] (original emphasis)

## Deflationist theories

There are a range of theories that call into question the fundamental premise of robust theories that truth does in fact have a substantial nature which needs to be explained. Ramsey (1927) held that the concept of truth was essentially redundant. In saying “it is true that snow is white” we are actually adding nothing to saying “snow is white”. The latter assumes or presumes the idea of truth and there nothing else to be said.

Strawson (1950) held that truth was essentially performative in that, in saying “it is true that snow is white” we are really just recommending or agreeing to the claim, so the truth predicate is not a property but an endorsement.

Quine (1992) argued that truth was disquotational. That is:

“’Snow is white is true’ if and only if snow is white. To ascribe truth to the sentence is to ascribe whiteness to snow; such is the correspondence in this example. Ascription of truth just cancels the quotation marks. Truth is disquotation.” (Quine 1992, p. 78)

Horwich (1991) held what he called a minimalist theory of truth. This has no theory of what truth is, but says simply that it is a logical system that has as its axioms every single instance of the general propositional form “The proposition that *p* is true iff *p*”. There will be an infinite number of these, for example “the proposition that *snow is white* is true if and only if *snow is white*”.

As can be seen, each of these theories, in different ways, denies that there should be a substantive explanation of the concept of truth. This is, in general, not a conclusion that many philosophers accept, and there are particular criticisms of each of the individual approaches - see  Lynch (2001) Section VI for details.